# The Gibbs Representation of 3D Rotations

*I.R. Peterson*

Centre for Molecular and Biomolecular Electronics,

Coventry University, NES, Priory Street, Coventry, CV1 5FB, UK

## *Abstract*

This paper revisits the little-known Gibbs-Rodrigues representation of rotations in a three-dimensional space and demonstrates a set of algorithms for handling it. In this representation the rotation is itself represented as a three-dimensional vector. The vector is parallel to the axis of rotation and its three components transform covariantly on change of coordinates. The mapping from rotations to vectors is 1:1 apart from computation error. The discontinuities of the representation require special handling but are not problematic. The rotation matrix can be generated efficiently from the vector without the use of transcendental functions, and vice-versa. The representation is more efficient than Euler angles, has affinities with Hassenpflug's Argyris angles and is very closely related to the quaternion representation. While the quaternion representation avoids the discontinuities inherent in any 3-component representation, this problem is readily overcome. The present paper gives efficient algorithms for computing the set of rotations which map a given vector to another of the same length and the rotation which maps a given pair of vectors to another pair of the same length and subtended angle.

Email: i.peterson@coventry.ac.uk Tel: +44 24 7688.8376 Fax: +44 24 7688.8702

# Introduction

When handling representations of three-dimensional space in a computer, there are many occasions on which it is necessary to convert the positions of points from one coordinate system to another rotated relative to it, e.g. [1]. Rotations also describe the movement of rigid bodies [2,3], the orientation of e.g. amino and nucleic acid residues in biological polymers [4] and the attitude of e.g. robot-held tools [5]. Rotation is a linear transformation represented by a $3\times3$ matrix, but the vast majority of $3\times3$ matrices are not rotations. A matrix is only a valid rotation if it is unitary and has a determinant of +1 [6]. The space of rotation matrices has three degrees of freedom, i.e. it is three-dimensional, but is not a linear subspace of the 9-dimensional space of possible matrices. There is a requirement for an efficient way of encoding and decoding rotations as sets of three real numbers.

The most common current solution to this problem is the Euler representation [7,8,9,10,11]. There are three Euler angles $\theta$, $\phi$ and $\psi$, each of which is physically an angle of a rotation performed in sequence about a set of coordinate axes. While no doubt inspired by the usual decomposition of vectors into components parallel to the *x*-, *y*- and *z*-axes, there is a difference in that rotations about different axes do not commute. Permuting the order in which the axes are taken causes the Euler angles to change in a way which is not obvious [12].

The overall rotation matrix in the Euler representation depends on the sines and cosines of $\theta$, $\phi$ and $\psi$ and their products. The inverse operation of determining the Euler angles describing a given rotation matrix requires the evaluation of inverse trigonometric functions [13] in a procedure with many distinct cases. All of these functions are transcendental, typically requiring many elementary operations even if the most efficient table lookup routines are used [14,15].

The present paper revisits the Gibbs-Rodrigues representation [16,17]. It is a covariant and computationally-efficient way of representing a rotation with three values. Transformation of the coordinate system of the space causes the three values taken together to transform like the components of a vector, which is parallel to the rotation axis. Each rotation is represented by a unique vector, although there is a two-dimensional subset of discontinuities. Non-transcendental algorithms exist for transforming back and forth between the matrix and vector forms, and for determining the vector representing a rotation defined in terms of commonly-encountered geometrical constraints.

## A Linear Representation in *N* Dimensions

The representation of 3D rotations to be proposed here is based on the following well-known general transformation of a real rotation matrix $U$ of any size [6]:

$$S = (U - I)(U + I)^{-1} \quad \text{(Equation 1)}$$

where $I$ is the unit matrix. Since all functions of a given matrix commute amongst themselves, the order of the factors of Equation 1 may be inverted. The transformation is valid only for rotation matrices, because for those unitary matrices which cannot be continuously transformed by infinitesimal rotations into $I$, the matrix $U + I$ is always singular. It is readily verified that if $U$ is unitary, and the transformed matrix $S$ exists, then $S$ is antisymmetric. The original matrix may be regenerated from $S$ using the inverse transformation:

$$U = (I + S)(I - S)^{-1} \quad \text{(Equation 2)}$$

The advantage of this antisymmetric representation is that the space of such matrices is a $\frac{1}{2}N(N-1)$–dimensional linear subspace of the $N^2$–dimensional space of all $N \times N$ matrices. Valid antisymmetric matrices are readily generated and each is completely specified by its subdiagonal coefficients $S_{ij}$ with $i > j$.

It is well known that all the functions of a given matrix have the same right and left eigenvectors as the matrix itself. Since a real antisymmetric matrix is antihermitian, all the eigenvalues of an antisymmetric matrix $S$ are pure imaginary $i\sigma$. The corresponding eigenvalue of $U$ is $(1 + i\sigma)/(1 - i\sigma)$, which has unit modulus.

For any eigenvalue $i\sigma$, the set of eigenvalues also includes its negative $-i\sigma$. The eigenvectors are in general complex. The one eigenvalue $i\sigma$ has distinct right and left eigenvectors, which are however complex conjugates $\vec{v} \pm i\vec{w}$. The right and left eigenvectors of $-i\sigma$ are the same two taken in reverse order $\vec{v} \mp i\vec{w}$ . If $\sigma$ is nonzero, the real vectors $\vec{v}$ and $\vec{w}$ corresponding to the pair of eigenvalues $\pm i\sigma$ are orthogonal and of equal length, which can be chosen to be unity [6].

In spaces of odd dimensionality including 3D, at least one of the eigenvalues must be its own negative, i.e. it must be zero. The corresponding nullvector $\vec{u}$ (both right and left) is then real, and can be seen from Equation 1 to be an axis of the rotation matrix $U$ . The complete set of nullvectors $\vec{u}$ and real pairs $\left(\vec{v}, \vec{w}\right)$ forms a real basis for the whole vector space.

## The Vector Representation in 3 Dimensions

The algorithm of Equation 1 requires the inversion of an $N \times N$ matrix, which can be performed in $O\left(N^4\right)$ elementary operations. It is possible to improve the computational efficiency of the antisymmetric representation by taking advantage of the particular properties of real antisymmetric matrices in 3D.

In three dimensions, $S$ has three independent components, and can be expressed in terms of a vector $\vec{r}$ , using the notation of Cartesian tensors [18]:

$$S_{ij} = \varepsilon_{ijk} r_k \quad \text{(Equation 3)}$$

$\varepsilon_{ijk}$ is the unit antisymmetric tensor, and the Einstein convention implies summation over the repeated suffix $k$ . Equation 3 can be inverted to express the vector $\vec{r}$ in terms of $S$ :

$$r_i = \tfrac{1}{2} \varepsilon_{ijk} S_{jk} \quad \text{(Equation 4)}$$

In the coordinate system defined by its eigenvectors, $S$ can be expressed as follows [6]:

$$S = \begin{bmatrix} 0 & 0 & 0 \\ 0 & 0 & \sigma \\ 0 & -\sigma & 0 \end{bmatrix} \quad \text{(Equation 5)}$$

from which it readily follows from Equation 4 that:

$$r_i = [\sigma \ 0 \ 0]_i = \sigma u_i \qquad \text{(Equation 6)}$$

Hence the vector $\vec{r}$ is a null-vector of $S$, parallel to the axis of the rotation $\vec{u}$, and its length determines the nontrivial eigenvalues $\pm i\sigma$ of $S$. With the help of Equation 2, the rotation matrix $U$ may also be expressed in this basis :

$$U = \frac{1}{1+\sigma^2}\begin{bmatrix} 1+\sigma^2 & 0 & 0 \\ 0 & 1-\sigma^2 & 2\sigma \\ 0 & -2\sigma & 1-\sigma^2 \end{bmatrix} \qquad \text{(Equation 7)}$$

Not only is the vector $\vec{r}$ parallel to the axis of the rotation, but it can be seen from Equation 7 that its length $\sigma$ is related to the angle of the rotation:

$$\sigma = \tan(\theta/2) \qquad \text{(Equation 8)}$$

The function $\tan(\theta/2)$ has periodicity $2\pi$, so that the uniqueness of $\sigma$ is consistent with the multivalued nature of $\theta$.

Equation 7 indicates that in 3D there is a algorithm simpler than Equation 2 for computing $U$. The three covariant matrices $\varepsilon_{ijk}u_k$, $u_i u_j$ and $\delta_{ij}$ are linearly independent and have the same left and right eigenvectors as $U$. Hence it is possible to express $U$ as a linear combination of them, which is readily seen to be:

$$U_{ij} = \frac{\left(1-\sigma^2\right)\delta_{ij} + 2\sigma^2 u_i u_j + 2\sigma\,\varepsilon_{ijk}u_k}{1+\sigma^2} \qquad \text{(Equation 9)}$$

The parameter $\sigma$ in Equation 5 to Equation 9 can be determined by taking the square root of $|\vec{r}|^2 = r_k r_k$. However there is no need separately to determine $\sigma$ or $\vec{u}$, because Equation 9 is readily re-expressed directly in terms of the components $r_i$ of the vector $\vec{r}$ :

$$U_{ij} = \frac{(1 - r_k r_k)\delta_{ij} + 2r_i r_j + 2\varepsilon_{ijk} r_k}{1 + r_k r_k}$$

. . . . . . . . . . . . . . . . . . . . . . . . (Equation 10)

This expression can be evaluated in $O(N^2)$ of the four elementary operations of addition, subtraction, multiplication and division which are implemented in the hardware of most modern computers.

Equation 10 expressing a rotation matrix in terms of a vector justifies a use of language in which the vector $\vec{r}$ "rotates" another vector. This operation combining two vectors, which may be symbolized ®, is neither associative nor commutative, but it does have the symmetry that the inverse operation is performed by the vector $-\vec{r}$ :

$$\left(-\vec{r}\right) ® \left(\vec{r} ® \vec{s}\right) \equiv \vec{s}$$ . . . . . . . . . . . . . . . . . . . . . . . . (Equation 11)

In this representation, a rotation whose angle is within machine accuracy equal to $\pi$ must be treated as an exception. According to Equation 8, its vector representation has infinite length. Although such rotations only occur with very low probability, it is essential to represent them and they can occur around any axis. Since a rotation of $\pi$ is nontrivial yet its own inverse, a discontinuity is inevitable in any unique representation with the desirable property of Equation 11. However the problem is readily managed if, as is most probable, the computations are performed with floating-point numbers. In Equation 1, a rotation angle of $\pi$ is signalled by the singularity of the denominator $U + I$ , which is then seen from Equation 10 to be equal to $2u_i u_j$ . From this it is a simple matter to determine $u_i$ . The vector $\vec{r}$ is then calculated from Equation 6, by approximating to $\sigma$ with the largest floating point number available to the machine. This approximation to $\sigma$ may be gross, but the approximation to the angle of rotation $\theta$ it represents is excellent. When using this strategy, the implementation of Equation 10 must handle such large numbers correctly, but the error in the regeneration of the matrix $U$ is typically much less than the computational error at other values of angle. Hence, as long as the exceptions are correctly handled, the transformation from vector to matrix and back is well-conditioned, with an error no greater than that of the elementary operations.

It should be noted that the Gibbs representation of 3D rotations has affinities with that proposed by Argyris and Hassenpflug [19, 20, 21]. The latter representation differs in that the length of the vector is equal to the rotation angle. The matrix functions which in the Argyris representation correspond to Equation 1 and Equation 2 and which link unit modulus $\exp(i\theta)$ and pure imaginary $i\theta$ eigenvalues are the logarithm and the exponential, respectively. Like the rotation angle itself, the Argyris angles never become infinite but are not unique. They also display the symmetry of Equation 11 and hence the principal value has a discontinuity for angles of $\pi$. This representation is less computationally efficient than the present proposal, as the relationship between the vector and matrix involves transcendental functions.

The present proposal also has close connections to the quaternion representation of rotations [22,23,24], and in fact the vector in the present representation is the ratio of the imaginary to real parts of the corresponding quaternion. In the quaternion representation, the discontinuity mentioned above at rotation angles of $\pi$ transforms harmlessly into the change of sign of the real part, and there is no need to handle any exceptions. However it is slightly less efficient than the present representation, requiring more elementary operations and regular extraction of square roots. Additionally, it is less accessible to intuition. In the next Section, vector representation algorithms will be presented whose equivalent quaternion form appears not to be known.

## Manipulations

The value of this representation will be judged by the ease with which it allows rotational computations to be performed. In fact, there are a number of calculations which become significantly easier than in the Euler, Hassenpflug or even the quaternion representations.

A solution to the problem of determining the vector representing a rotation given the matrix of the latter has already been given in Equation 2 and Equation 4. However just as in 3D it is possible to convert Equation 1 into the much more rapidly evaluated Equation 10, so there exists a more efficient 3D form for its inverse. Since the matrix has 9 elements while the vector has only 3, this problem is highly overdetermined and there are many equivalent forms of the solution. The most elegant, explicitly covariant and readily verified by back substitution in Equation 10, is the following:

$$r_i = \frac{\varepsilon_{ijk} U_{jk}}{1 + U_{kk}} \qquad \text{(Equation 12)}$$

As for Equation 10, the evaluation of this expression requires only $O(N^2)$ of the four elementary operations. The only exception is division by zero, to which a meaningful result preserving computational accuracy can nevertheless be attached:

$$r_i = L.\left(\frac{\delta_{iK} + U_{iK}}{1 + U_{KK}}\right) \qquad \text{(Equation 13)}$$

where $L$ is the largest floating-point number of the machine and $K$ is the row and column in which the largest diagonal element of $U$, $U_{KK}$, occurs. Equation 12 and Equation 13 together are markedly simpler than the corresponding algorithm in the Euler representation for determining the angles θ, ϕ and ψ.

A second common problem is to determine the set of rotations $r_i$ which map the vector $p_i$ to the vector $q_i$. Since rotation preserves length, both $p_i$ and $q_i$ must have equal length, which can be taken without significant loss of generality to be unity.

Rotation of a vector $p_i$ about an axis parallel to the unit vector $\xi_i$ preserves not only the length of $p_i$ but also the angle it subtends to $\xi_i$. The set of all vectors subtending the same angle to both $p_i$ and $q_i$ is clearly the perpendicular bisector to the line joining $p_i$ and $q_i$. This is the plane through the origin perpendicular to $p_i - q_i$, and a convenient basis is the pair $p_i + q_i$ and $\varepsilon_{ijk} p_j q_k$.

The calculation of the angle of rotation around an arbitrary axis in this plane parallel to $\xi_i$ is simplified by symbolizing as $\alpha$ the angle between $p_i$ and the perpendicular bisector, and as $\beta$ the angle between $p_i + q_i$ and $\xi_i$. Then, in an obvious orthonormal coordinate system:

$$p_i = [\cos\alpha\cos\beta, \quad \cos\alpha\sin\beta, \quad \sin\alpha]_i$$
$$q_i = [\cos\alpha\cos\beta, \quad \cos\alpha\sin\beta, \quad -\sin\alpha]_i \quad \text{.......... (Equation 14)}$$
$$\xi_i = [1, \quad 0, \quad 0]_i$$

Clearly, the angle $\theta$ of the rotation around $\xi_i$ which maps $p_i$ onto $q_i$ is given by:

$$\tan\left(\frac{\theta}{2}\right) = \frac{\sin\alpha}{\cos\alpha\sin\beta} \quad \text{........................ (Equation 15)}$$

Further calculation gives the vector $r_i \equiv \xi_i \tan(\theta/2)$ representing the rotation in simplest form as:

$$r_i = \frac{\varepsilon_{ijk} p_j q_k + \gamma(p_i + q_i)}{1 + p_k q_k} \quad \text{...................... (Equation 16)}$$

$\gamma$ depends on both $\alpha$ and $\beta$. However $\beta$ is arbitrary, and since the orientation between the axis and any other direction is readily determined directly from $r_i$, it is not necessary to give this relationship explicitly. Equation 16 generates the set of vectors that rotate $p_i$ to $q_i$. The set forms a straight line, and for any vector in the set it is immediately possible to state what the axis of the rotation is, and the angle of rotation about it which maps $p_i$ to $q_i$.

For non-unit vectors, the formula of Equation 16 is readily modified to give:

$$r_i = \frac{\varepsilon_{ijk} p_j q_k + \gamma(p_i + q_i)}{p_k (p_k + q_k)} \quad \text{...................... (Equation 17)}$$

The solution of Equation 17 can be applied to the problem of determining the vector which rotates a pair of vectors $p_{1i}$ and $p_{2i}$ to a second pair $q_{1i}$ and $q_{2i}$. It is the point on the line of vectors which rotate $p_{1i}$ to $q_{1i}$ which is orthogonal to $p_{2i} - q_{2i}$, i.e.:

$$r_i = \frac{\varepsilon_{ijk} p_{1j} q_{1k} + \gamma(p_{1i} + q_{1i})}{p_{1k}(p_{1k} + q_{1k})}$$
$$\gamma = -\frac{\varepsilon_{ijk} p_{1j} q_{1k} (p_{2i} - q_{2i})}{(p_{1i} + q_{1i})(p_{2i} - q_{2i})} \quad \text{...................... (Equation 18)}$$

Within the present representation, the above algorithm gives an elegant solution to the problem solved in the quaternion representation by the “belt trick” [23]. When generating a shape by sweeping a closed loop along a curve, the rotation of the closed loop can be chosen to map directions fixed to the loop onto the tangent and curvature vectors of the curve. It should be noted that Equation 18 produces a result for essentially all sets of four unit vectors. It will only be a valid solution if the appropriate lengths and subtended angles of the two pairs are equal.

In some situations it may be desirable to calculate the overall rotation resulting from the action of two individual vectors without converting first to matrix form. Successive application twice of Equation 10 and once of Equation 12 followed by extensive simplification leads to:

$$(r \otimes s)_i = \frac{r_i + s_i - \varepsilon_{ijk} r_j s_k}{1 - r_k s_k} \qquad \text{(Equation 19)}$$

where $\otimes$ represents the operation of successive rotation. This equation is clearly closely related to the formula for quaternion multiplication.

# Summary

This paper has demonstrated that the Gibbs representation of 3-dimensional rotations has significant advantages over other current methods of representation. It is very closely related to the quaternion representation, whose freedom from discontinuities is often considered an advantage. However the latter are readily handled computationally. The Gibbs representation allows rapid conversion to and from the matrix representation, simplifies a number of naturally-arising calculations, and since it can be visualised allows greater intuition. For convenience the main formulae of this representation (apart from exception handling) are listed here.

## Conversion from Vector to Matrix

$$U_{ij} = \frac{(1 - r_k r_k)\delta_{ij} + 2 r_i r_j + 2\varepsilon_{ijk} r_k}{1 + r_k r_k} \qquad \text{(Equation 10)}$$

## Conversion from Matrix to Vector

$$r_i = \frac{\varepsilon_{ijk} U_{jk}}{1 + U_{kk}} \qquad \text{. . . . . . . . . . . . . . . . . . . . . . . . . . . . . (Equation 12)}$$

## Set rotating one vector to another

$$r_i = \frac{\varepsilon_{ijk} p_j q_k + \gamma \left( p_i + q_i \right)}{p_k \left( p_k + q_k \right)} \qquad \text{. . . . . . . . . . . . . . . . . . . . . . . . .(Equation 17)}$$

## Vector rotating one pair to another

$$\begin{aligned} r_i &= \frac{\varepsilon_{ijk} p_{1j} q_{1k} + \gamma \left( p_{1i} + q_{1i} \right)}{p_{1k} \left( p_{1k} + q_{1k} \right)} \\ \gamma &= -\frac{\varepsilon_{ijk} p_{1j} q_{1k} \left( p_{2i} - q_{2i} \right)}{\left( p_{1i} + q_{1i} \right) \left( p_{2i} - q_{2i} \right)} \end{aligned} \qquad \text{. . . . . . . . . . . . . . . . . . . . . . . (Equation 18)}$$

## Sequence of two rotations

$$\left( r \otimes s \right)_i = \frac{r_i + s_i - \varepsilon_{ijk} r_j s_k}{1 - r_k s_k} \qquad \text{. . . . . . . (Equation 19)}$$

# LIST OF CAPTIONS